# Oscillating Decay Rate in Electron Capture and the Neutrino Mass Difference

Murray Peshkin*
*Physics Division, Argonne National Laboratory, Argonne, Illinois 60439, USA*

ABSTRACT

Reported oscillations in the rate of decay of certain ions by K-electron capture have raised questions about whether and how such oscillations can arise in quantum mechanical theory and whether they can measure the neutrino mass difference. Here I show that simple principles of quantum mechanics answer some questions and clarify what must be done theoretically or experimentally to answer some others. The principal result is that quantum mechanics does allow mass-difference-dependent oscillations in principle, but it imposes conditions not obeyed by the approximate dynamical models that have been put forth up to now. What needs to be done experimentally and theoretically is discussed.

## I. INTRODUCTON

It has been reported [1,2] that the decay rate $R(t)$ observed in certain electron-capture experiments differs from the usual exponential law, being described better by

$$R(t) \propto e^{-\lambda t}\left(1 + a\cos(\omega t + \phi)\right) \qquad (1.1)$$

with non-vanishing $a$. The time $t$ was measured from the time of a nuclear collision in which the decaying system was created. In the most refined experiment [2], hydrogen-like ions consisting of a $^{142}\text{Pm}_{61}$ nucleus and one electron in the K shell decayed by electron capture leaving a two-body final state containing a bare $^{142}\text{Nd}_{60}$ nucleus and a neutrino. The reported best-fit values of the parameters were $\lambda = 0.013 \text{ sec}^{-1}$, $\omega = 0.88 \text{ sec}^{-1}$, $\varphi = 2.4$ rad, and $a=0.107(24)$. Hydrogen-like $^{142}\text{Pm}$ also decays by positron emission. In that channel, the value of the oscillation amplitude that best fit the data was $a=0.027(27)$, indistinguishable from zero and significantly smaller than the amplitude in the electron-capture channel. The measured value of $\lambda$ was consistent with that in the electron-capture channel. All measurements were carried out in the GSI storage ring so that the decay took place in the presence of strong magnetic fields, stochastic-cooling and electron-cooling interactions, and interactions with the detectors used to track the orbital period of the ions. Although the current experimental results have yet to be fully confirmed, they raise questions that may apply usefully to a future experiment of the same kind, or possibly of a related kind.

The origin of these "GSI oscillations", as they have come to be called, is not now known. A possibility that has been considered and eliminated is based on hyperfine state mixing in the parent $^{142}\text{Pm}$ ion. The nucleus has spin 1. Of the two hyperfine states with total angular momentum $F$ equal to 1/2 and 3/2, only $F=1/2$ can decay by a Gamow-Teller transition into a state containing just a neutrino and a spin-zero daughter nucleus. The magnetic field in the storage ring mixes the two states to create in effect an oscillating source for the decay. However it has been shown [3,4] that such oscillations in the magnetic field of a storage ring would be about ten orders of magnitude too rapid to account for the reported experimental oscillations.

This paper addresses a second approach that has been discussed in the literature [5-9]. The final state after electron capture contains two channels, one for each of two neutrino mass states. (Strictly, there are three channels for three mass states but two suffice for present purposes.) The partial decay rates for the two are nearly equal because the neutrino mass difference is tiny in comparison with the decay energy. However, interference between amplitudes dependent upon the two masses could lead to oscillations dependent upon the mass difference. Then the observed oscillation frequency could provide an independent measurement of the neutrino mass difference. Heuristic models [5-7,9] have made use of the interactions between the ions and external fields or the apparatus used to detect the decay to carry out that program, with different results depending upon their assumptions. Those models have been refuted by a first-principles argument [8] based on the idea that the interference assumed in those models is forbidden by the orthogonality of the neutrino mass eigenstates. It is shown below that the challenge [8] does in fact negate the existing models but not necessarily all possible models.

The law of exponential decay into a single channel is widely successful in describing experiments but notoriously deficient in its derivation from quantum mechanical principles [10-12]. Interference effects between decaying states can be delicate and need to be treated as nearly rigorously as possible. Here I address two questions raised by that challenge:

1. Does quantum mechanics allow oscillations in the electron capture rate to arise from the mass difference? If so, can the decay by positron emission fail to have similar oscillations? In fact, why do we not then see oscillations in any of the many other cases of nuclear decay into competing channels?
2. What limitations does quantum mechanics place on candidate models relating oscillating decay to the neutrino mass difference and what can be done experimentally to overcome the theoretical uncertainties?

Section II below addresses the first of these questions and Section III the second. My conclusions are summarized in Section IV.

## II. INTERFERENCE, MASS DIFFERENCE, AND OSCILLATIONS

The Hilbert space under consideration contains four mutually orthogonal channels: one (called $p$ below) with a parent ion present, one (called 1) with a daughter nucleus and a neutrino with mass $m_1$, one (called 2) with the same daughter and a neutrino of mass $m_2$, and one (called +) with a positron, a different daughter ion, and a neutrino of either mass. In each channel, the dynamical variables are the position or momentum of each particle present and whatever spins or other internal variables apply, plus the variables of electrons used to cool the ion beam and of the stochastic cooling apparatus, and of the Schottky detectors used to find the time of decay. In reality there are at least two additional channels, the one mentioned above containing a third neutrino whose inclusion would change nothing important below, and one containing the nuclei and other particles that were present when the parent ion was born. Neglecting the second amounts to the assumption that there is a time $t=0$ when, to an adequate approximation, only the parent ion is present. Then the wave function $\psi(t)$ for the entire system can be written as

$$\psi(t) = \varphi_p(t) + \varphi_1(t) + \varphi_2(t) + \varphi_+(t), \tag{2.1}$$

where

$$\varphi_j(t) = P_j \psi(t) \tag{2.2}$$

The $P_j$ are the projectors on the four channels and the $\varphi_j$ are functions of the dynamical variables of those channels, which include the above-mentioned external variables. By assumption

$$\varphi_1(0) = \varphi_2(0) = \varphi_+(0) = 0. \tag{2.3}$$

The Hamiltonian has the form

$$H = H_0 + V, \tag{2.4}$$

where $V$ is the weak interaction and $H_0$ is everything else in $H$.

$$H_0 = P_p H P_p + P_1 H P_1 + P_2 H P_2 + P_+ H P_+ = H_p + H_1 + H_2 + H_+ \tag{2.5}$$

$$\begin{aligned} V &= P_p H P_1 + P_1 H P_p + P_p H P_2 + P_2 H P_p + P_p H P_+ + P_+ H P_p \\ &= V_{p1} + V_{1p} + V_{p2} + V_{2p} + V_{p+} + V_{+p} \end{aligned} \tag{2.6}$$

From the Schroedinger equation or its relativistic generalization,

$$i\dot{\varphi}_p(t) = H_p \varphi_p(t) + V_{p1} \varphi_1(t) + V_{p2} \varphi_2(t) + V_{p+} \varphi_+(t) \tag{2.7}$$
$$i\dot{\varphi}_1(t) = H_1 \varphi_1(t) + V_{1p} \varphi_p(t) \tag{2.8}$$
$$i\dot{\varphi}_2(t) = H_2 \varphi_2(t) + V_{2p} \varphi_p(t) \tag{2.9}$$
$$i\dot{\varphi}_+(t) = H_+ \varphi_+(t) + V_{+p} \varphi_p(t). \tag{2.10}$$

The survival probability $S(t)$ of the parent state and its rate of decay $R(t)$ are given by

$$S(t) = \langle \varphi_p(t) | \varphi_p(t) \rangle \tag{2.11}$$

$$R(t) = -\frac{dS}{dt} = +\frac{d}{dt}\left[\langle \varphi_1(t) | \varphi_1(t) \rangle + \langle \varphi_2(t) | \varphi_2(t) \rangle + \langle \varphi_+(t) | \varphi_+(t) \rangle\right] \tag{2.12}$$

In Eqs.(2.11, 2,12) and hereafter, the inner products include the trace of a density matrix involving the external variables. Those equations remain valid if the external variables are entangled with those of the daughter ion and the neutrino in $\varphi_1$ and $\varphi_2$; *i.e.* if the density matrix implied by $\varphi_1$ is different from that implied by $\varphi_2$. Eq.(2.12) shows that any oscillations in the decay rate cannot arise from interference between the wave functions in channel 1 and channel 2. (Decay oscillations are in this respect very different from the familiar spatial oscillations of solar or reactor neutrinos; those are measured by the expectation of a projector of the e-neutrino, which is off-diagonal in the mass basis.)

That quantum mechanics rigorously excludes the possibility of decay oscillations that arise from interference between the two mass channels was first noted by Flambaum [6]. As seen explicitly above, it applies in both the presence and the absence of external interactions and independently even of the unlikely possibility that the decay interaction *V* itself depends upon external influences. However, that reasoning does not exclude the possibility of interference between terms depending upon the two mass values in each of the two mass-channel wave functions; the unavoidable indirect coupling of $\varphi_1$ and $\varphi_2$ through the direct coupling of each to $\varphi_p$ can in principle induce oscillations proportional to the neutrino mass difference. From Eqs.(2.7, 2.8),

$$i\ddot{\varphi}_1 = \dot{H}_1\varphi_1 + H_1\dot{\varphi}_1 - iV_{1p}\left[H_p\varphi_p + V_{p1}\varphi_1 + V_{p2}\varphi_2 + V_{p+}\varphi_+\right] \tag{2.13}$$

From Eq.(2.9), $\varphi_2$ depends upon $m_2$ through $H_2$. Therefore from Eq.(2.13), $\varphi_1$ depends upon both masses, and similarly for $\varphi_2$. Interference between terms depending upon the two masses may appear within $\langle \varphi_1(t) | \varphi_1(t) \rangle$ and $\langle \varphi_2(t) | \varphi_2(t) \rangle$ in Eq.(2.12), and therefore also in the decay rate. Those interference terms are not necessarily small. The weak interaction *V* appears quadratically in Eq.(2.13), but the times of importance are comparable with the halflife, which is proportional to $\hbar/V$, or in our units to $1/V$. The mixing, like exponential decay itself, is not correctly described as a low order perturbation. The decay rate $R(t)$ rises from 0 at $t=0$ before becoming exponential [12]. That cannot come out of a perturbative approximate solution of Eq.(2.13) containing only low powers of *V*. In other language, the amplitudes for finding the ion in its parent state at times comparable with the halflife contain important contributions from Feynman diagrams containing multiple loops in which a daughter nucleus and a neutrino are present, and those loops involve both masses.

To this point it has been shown only that quantum mechanical principle cannot alone exclude interference effects in multi-channel decay. A stronger statement can be made in the case of electron capture by introducing the flavor basis and considering the limiting case where $m_1 - m_2 \to 0$.

$$\varphi_e(t) = [\varphi_1(t) + \varphi_2(t)]/\sqrt{2}$$
$$\varphi_\mu(t) = [\varphi_1(t) - \varphi_2(t)]/\sqrt{2}$$
(2.14)

Electron capture couples the parent ion only to electron neutrinos. From Eqs.(2.8, 2.9), $\varphi_1(t) = \varphi_2(t)$ for all $t$ in the limit $m_1 - m_2 \to 0$. Then for small $m_1 - m_2$, $\varphi_1(t)$ and $\varphi_2(t)$ must be nearly equal, depending upon both masses and adding constructively in $\varphi_e(t)$. In Eq.(2.13), the product $V_{1p}V_{p2}$, which is the source of the interference, is nearly equal to the product $V_{1p}V_{p1}$, which is the source of the decay without interference between the two mass values.

The presence of interference between terms dependent upon the two masses in channels 1 and 2 enables, but does not require non-negligible decay oscillations as a matter of quantum mechanical principle. Whether such oscillations are implied by the interference, and if so exactly how they depend upon the mass difference can be answered only by a dynamical model. No satisfactory candidate model appears to have has been offered at present, but with or without such a dynamical model the question arises as to whether the presence of non-negligible oscillations in the decay by electron capture implies oscillations of a similar magnitude in the decay by positron emission. It does not. The oscillating parent state is the source of the positron emission, as is seen in Eq.(2.10), so some oscillation in the positron emission must exist, but it can be weak for several reasons. For instance, the natural frequencies in $H_+$ may be much lower than the oscillation frequency. Also, the final state in positron emission is a three-body state so that $\varphi_+(t)$ has subchannels. The decay into each of those may oscillate, but not necessarily in phase with others, so that any oscillation in the total rate of decay by positron emission may be negligible.

The unique feature of the coupling between the two mass channels in electron capture, that $\varphi_1$ and $\varphi_2$ are nearly equal, as are $V_{p1}$ and $V_{p2}$ in Eq.(2.13) because only an electron neutrino is created in the capture process, has no counterpart in the general case of two-channel decays. Therefore there is no reason to expect non-negligible oscillations in ordinary atomic or nuclear exponential decay phenomena where no such symmetry exists. That applies to positron emission in $P_m$ decay as well.

Absent a satisfactory dynamical model, the question of whether interference based on unequal masses necessitates oscillations in decay is currently unanswered. A possible affirmative hint has been provided [12] by a two-channel generalization of Winter's model [13,14]. In that numerically solvable model, a nonrelativistic particle moving in one dimension is acted upon by a delta-function potential analogous to the confining potential in alpha-particle emission. In the generalization, the two channels have different masses and the delta function potential mixes them in analogy to the mixing of neutrino mass states in weak decay. For some values of its parameters, the model does produce oscillations in the decay rate on time scales comparable with the halflife. That simple model is far from being a realistic representation of neutrino emission but it does have some suggestive features in common with reality.

## III. THEORETICAL CONSTRAINTS, EXTERNAL INTERACTIONS, AND NEEDED EXPERIMENTS

In the most general case, the parent state survival probability is given by

$$S(t) = \langle P_p \psi(t) | P_p \psi(t) \rangle = \langle \psi(0) | e^{iHt} P_p e^{-iHt} | \psi(0) \rangle$$
$$= \iint ds \, ds' \iint d^3\mathbf{K} \, d^3\mathbf{K}' \langle \psi(0) | \mathbf{K}, s \rangle \langle \mathbf{K}, s | e^{iHt} P_p e^{-iHt} | \mathbf{K}', s' \rangle \langle \mathbf{K}', s' | \psi(0) \rangle \quad (3.1)$$

Here **K** stands for the momentum of the parent ion, and $s$ for its internal and spin variables, all of them invariant under translation.

$$\left[ \hat{\mathbf{K}}, \hat{s} \right] = \left[ \hat{\mathbf{K}}, P_p \right] = 0 \quad (3.2)$$

In reality, the parent and daughter ion are subjected to external interactions. Nevertheless it is useful to consider what would happen in the absence of such interactions. In that case, $\hat{\mathbf{K}}$ also commutes with $H$ and

$$S(t) = \iint ds \, ds' \int d^3\mathbf{K} \langle \psi(0) | \mathbf{K}, s \rangle \langle \mathbf{K}, s | e^{iHt} P_p e^{-iHt} | \mathbf{K}, s' \rangle \langle \mathbf{K}, s' | \psi(0) \rangle \quad (3.3)$$

Eq.(3.3) shows that any model which finds a contribution to $S(t)$ from interference between different values of the parent ion's momentum independently of external interactions must have relied upon assumptions or approximations that are inconsistent with quantum mechanics. That is the case even if the interference arises separately in the two neutrino mass channels, as is demanded by the considerations in Section II above.

The necessity for external interactions to influence a possible connection between decay oscillations and the neutrino mass difference raises the question as to whether such oscillations are not then impossible. For example, the interaction with the cooling electrons manifestly alters the motion of the ions. Will interference effects be wiped out or at least randomized by the ion's wave function losing its phase correlations? That question can be answered decisively only in the context of a dynamical model, but a reasonable argument can be given to show by analogy that no such problem will arise. Consider a spin ½ ion moving through a beam of spinless particles with at most a very weak spin-orbit interaction. Let the ion be polarized so that $\sigma_z = +1$. the ions will emerge the beam will emerge with $\sigma_z$ nearly equal to +1. The incident ion beam could also be thought of as a coherent mixture of $\sigma_x = +1$ and $\sigma_x = -1$. The phase relation between the two will not be lost, however strong the spin-independent interaction with the spinless particles. In the neutrino case, where the mass basis is analogous to the $\sigma_x$ basis, the relative phase of the two mass states should be unchanged by the interaction as long as the interaction itself is only weakly dependent upon the neutrino mass.

## IV. CONCLUSIONS

Quantum mechanics permits oscillations in the rate of decay by electron capture to arise from interference effects proportional to the neutrino mass difference. That comes about in consequence of the indirect coupling of two neutrino mass channels through their direct coupling to the decaying ion by the weak interaction. A contrary conclusion [6] resulted from the neglect of that coupling.

Quantum mechanical principles do not alone require non-negligible oscillations in the decay rate. A dynamical theory is required to infer decay oscillations from interferences in the channel wave functions and to relate any such oscillations to the neutrino mass difference.

Any oscillations due to interference between parts of the wave function dependent upon the two neutrino masses cannot arise from interference between the two mass channels. They must come about as the result of the appearance of both masses in the wave functions in each mass channel. That is a rigorous consequence of quantum mechanics even in the presence of entanglement with dynamical variables such as those of the Schottky detectors and the electron-cooling beam. A dynamical model that produces oscillations through interference between the two channels, as do all the currently proposed models of which I know [5,7-9], must contain assumptions forbidden by rigorous quantum mechanics.

The parent ion is the source of decay by positron emission as well as by electron capture. Eq.(2.10) implies that if the rate of decay of the parent ion oscillates, so must the wave function in the positron channel. However, even if the oscillations in the electron-capture decay rate are large, the rate of decay by positron emission should be expected to be negligible. Interference effects in the electron capture rate may be strong because the decay interaction creates only electron neutrinos, which contain the two mass states equally. No analogous symmetry relates positron emission to electron capture. For the same reason, oscillations are not to be expected in ordinary nuclear decay into two or more channels.

Given the complexity of the environment in a storage ring or even an ion trap, it may be difficult to include the effects of external fields and other interactions in a theoretical model, or even to justify neglecting those interactions on theoretical grounds. Then the possibility of measuring the neutrino mass difference by decay oscillations will have to rest on a theoretical model that ignores the external interactions and must be justified by experiments which show that the oscillations are independent of changes in the external interactions, including electron cooling beams and Schottky detectors. For that, an ion trap experiment would be especially advantageous if it should prove feasible.

Relativistic motion of the decaying ions cannot be an essential condition for oscillations. An ion nearly at rest would be just as good.

## ACKNOWLEDGMENTS

This material is based upon work supported by the U.S. Department of Energy, Office of Science, Office of Nuclear Physics, under contract number DE-AC02-06CH11357. I thank Avraham Gal, Boris J. Kayser, and Ernst Otten for valuable discussions.

=================

* email: peshkin@anl.gov